\documentclass[epj]{svjour}
\usepackage{graphicx}
\usepackage[referable]{threeparttablex}
\usepackage{subfigure}
\usepackage{amsmath}

\begin{document}
\title{Astrophysical signatures of leptonium}
\author{S.~C.~Ellis\inst{1,2}\thanks{sellis@aao.gov.au} \and J.~Bland-Hawthorn\inst{2}
%
}                     
\offprints{}          
\institute{Australian Astronomical Observatory, 105 Delhi Rd., North Ryde, NSW 2113, Australia \and
Sydney Institute for Astronomy, School of Physics, University of Sydney, NSW 2006, Australia}
\date{Received: date / Revised version: date}
%
\abstract{
More than $10^{43}$ positrons annihilate every second in the centre of our Galaxy yet, despite four decades of observations, their origin is still unknown.  Many candidates have been proposed, such as supernovae and low mass X-ray binaries.   However, these models are difficult to reconcile with the distribution of positrons, which are highly concentrated in the Galactic bulge, and therefore require specific propagation of the positrons through the interstellar medium.  Alternative sources include dark matter decay, or the supermassive black hole, both of which would have a naturally high bulge-to-disc ratio.
The chief difficulty in reconciling models with the observations is the intrinsically poor angular resolution of gamma-ray observations, which cannot resolve point sources. Essentially all of the positrons annihilate via the formation of positronium. This gives rise to the possibility of observing recombination lines of positronium emitted before the atom annihilates. These emission lines would be in the UV and the NIR, giving an increase in angular resolution of a factor of $10^{4}$ compared to gamma ray observations, and allowing the discrimination between point sources and truly diffuse emission.
Analogously to the formation of positronium, it is possible to form atoms of true muonium and true tauonium. Since muons and tauons are intrinsically unstable, the formation of such leptonium atoms will be localised to their places of origin. Thus observations of true muonium or true tauonium can provide another way to distinguish between truly diffuse sources such as dark matter decay, and an unresolved distribution of point sources.
} 
\maketitle

\section{Introduction}
\label{sec:hist}

The prediction\cite{dir28,dir30,dir31} and discovery\cite{and33} of positrons are rightly regarded as outstanding examples of the achievements of both theoretical and experimental physics.  One year after this discovery, Mohorovi{\v c}i{\'c}\cite{moh34} suggested the possibility of the existence of positronium (Ps), and suggested searching for astrophysical sources of Ps via its recombination lines.   This paper received little attention initially, and thereafter was forgotten for some time, and in fact the challenge of observing celestial recombination lines from Ps has still not been met, to which subject we shall return later.   However, during the development of quantum electrodynamics in the mid 1940s the idea of Ps atoms was predicted independently and separately by Pirenne \cite{pir44}, Ruark\cite{rua45}, Landau (unpublished work referred to in \cite{ali45}) and Wheeler\cite{whe46},  works that established its properties on a more secure theoretical footing.
Positronium was eventually discovered in 1951\cite{deu51} in laboratory experiments in which a beam of positrons were fired into cold gases. 

 Tantalising hints of astrophysical positronium were observed 
 in the late 1960s and early 1970s\cite{hay69,joh72}, which measured the gamma-ray spectrum of the Galactic centre, and found evidence for an emission line close to the electron-positron annihilation energy of 511~keV, but did not have sufficient energy resolution to provide an accurate identification.  Instead, the first identification  of celestial Ps came from  observations of solar flares\cite{chu73}, and  
in 1978 when the Galactic centre 511~keV line was also unambiguously identified\cite{lev78}. 
 Around the same time,  the first measurements of the Lyman~$\alpha$ recombination line of Ps were made  in laboratory experiments\cite{can75}.
 
Today, the state-of-the-art observations of Ps come from the SPI spectrometer onboard the INTEGRAL satellite.  Siegert et al.\cite{sie16} have performed the most recent analysis on 11 years of data covering the entire sky, which provides information on both the morphology of the radiation, and the spectrum of the radiation simultaneously, and a good review of Ps astrophysics is given by Prantzos et al.\cite{pra11}.  

Every second approximately $5 \times 10^{43}$ positrons annihilate with electrons in our Galaxy.  Almost all of these annihilate via the formation of Ps, as evidenced from  the annihilation spectrum,
 which shows both  two photon 511~keV line emission and a $< 511$~keV continuum emission.
 The positronium fraction can be determined from the ratio of the intensities of the line and continuum radiation\cite{pra11}, $I_{\rm 511}$ and $I_{\rm cont}$, since
 the two photon line emission arises from direct annihilation and from the  annihilation of the singlet state para-Ps, whereas the three photon continuum arises from the annihilation of the triplet state ortho-Ps, thus,
 \begin{equation}
 \label{eqn:fps}
 f_{\rm Ps} = \frac{8}{9\frac{I_{\rm 511}}{I_{\rm cont}}+6}.
 \end{equation}
The spectrum also displays a slightly broadened line, which together with the Ps fraction suggests that the annihilation is taking place primarily in a partially ionised interstellar medium (consisting of $\approx 70\%$ H, $\approx 28 \%$ He and $\approx 2\%$ heavier atoms, by mass) at $\approx 8000$~K\cite{chu05}.  This may be the result of positrons injected into a hot $\sim 10^{6}$~K gas,  which radiatively cools to these temperatures whereupon the annihilation time becomes less than the cooling time\cite{chu11}.

\begin{figure}
\centering \includegraphics[scale=0.6]{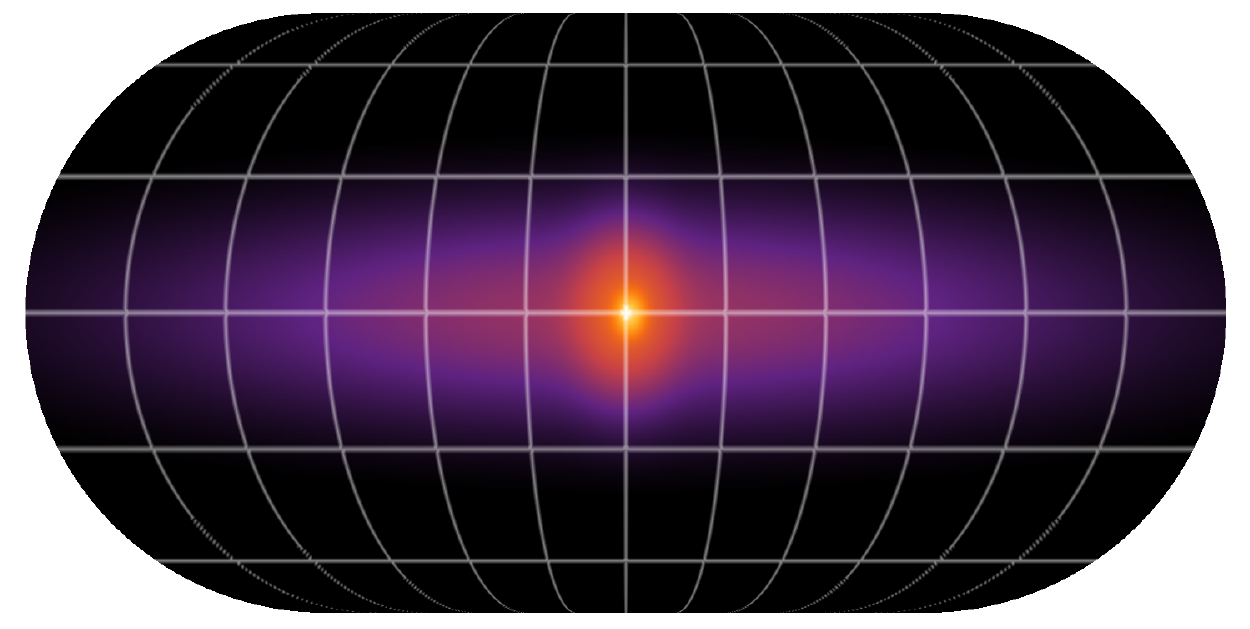}
\caption{An equal area Eckert IV projection of the cube root of the surface brightness of the best-fitting model image of the INTEGRAL SPI 511~keV emission, following Siegert et al.\protect\cite{sie16}.}
\label{fig:morph}
\end{figure}

The morphology of the 511~keV radiation is best described by a four component model\cite{ski14,sie16}, comprised of a narrow bulge, a broad bulge, a compact central source and a disc component.  It should be borne in mind that all analyses of the distribution of 511~keV radiation are necessarily model dependent, since images are reconstructed from a coded-mask detector; different authors may prefer physically motivated models, or simple parametric models.   For example, the disc emission was found to be asymmetric\cite{wei08}, but this asymmetry can also be explained by a small offset of the bulge emission combined with a symmetric disk\cite{bou10,sie16}.  Figure~\ref{fig:morph} shows the best fitting model components from Siegert et al.\cite{sie16}.

Reconciling the spectrum and morphology of the Ps annihilation radiation with astrophysical sources has been a challenge ever since the 511~keV radiation was first discovered.  Whilst there is no shortage of plausible candidates to produce the e$^{+}$, it is hard to accurately assess their importance.  In the first place there is no class of sources which are distributed like the 511~keV morphology, which has a very large bulge-disc ratio ($\approx 0.6$ in flux) and a rather broad disc.  However, since the positrons can propagate through the ISM before annihilation, this mismatch is not necessarily a problem if propagation distances are large ($\sim$~kpc).    In fact, the spectrum of annihilation radiation, and positron propagation models, both show that the e$^{+}$ annihilate  in warm partially ionised phases of the ISM.  Thus the morphology of the 511~keV radiation is actually a result of the  distribution of both the sources and the warm clouds in which the annihilation occurs.

A confounding factor in the difficulty of identifying the sources of positrons is the intrinsically low angular resolution of gamma-ray observatories.  INTEGRAL SPI has approximately 2 degree resolution, which is several times better than any of its predecessors, but which is nevertheless too coarse to resolve the emission.  
Despite this, Siegert et al.\cite{sie16b} recently discovered 511~keV radiation from a micro-quasar V404 Cygni during its 2015 outburst.  The temporal behaviour of the event, which showed a correlation between the 511~keV annihilation line emission and the 100 -- 200~keV hard X-ray continuum emission, allowed an unambiguous identification despite the lack of sensitivity to point sources.

In the next section (\S~\ref{sec:models}), we briefly review the current status of astrophysical models of positrons sources and propagation and their ability to account for the observed gamma-ray emission.  Thereafter, we come to the main theme of the paper in which we
 review  signatures other than the annihilation of e$^{+}$ and Ps, and their potential for resolving the identity of the sources of Galactic positrons.  In section~\ref{sec:recomb} we review the prospect of 
observing Ps recombination lines and the related possibility of observing spin-flip transitions.  Then in section~\ref{sec:lep} we review the potential information available from the analogous atoms of true muonium and true tauonium, which consist of bound states of $\mu$-$\bar{\mu}$ and $\tau$-$\bar{\tau}$ respectively\footnote{The suffix `onium' generally refers to an atom consisting of a particle - antiparticle pair.  However, in the case of muonium the name refers instead to an atom consisting of an antimuon and an electron, hence the somewhat awkward nomenclature `true muonium', and by analogy `true tauonium'}.  We summarise our results in section~\ref{sec:summ}.

\section{Astrophysical modelling of positron sources}
\label{sec:models}

Positrons may be produced by the following processes\cite{gue05},
\begin{eqnarray}
p & \rightarrow & n + {\rm e}^{+} + \nu_{\rm e} \\
\gamma + \gamma &\rightarrow& {\rm e}^{+} + {\rm e}^{-} \\
\pi^{+} & \rightarrow & \mu^{+} + \nu_{\mu}  \\
\mu^{+} & \rightarrow & {\rm e}^{+} + \nu_{\rm e} + \bar{\nu}_{\mu},
\end{eqnarray}
all of which occur in astrophysical processes.  Indeed, there is no shortage of potential sources of Galactic positrons.   Table~\ref{tab:cand} is taken from the excellent review by Prantzos et al.\cite{pra11}, and summarises the main properties of the candidate positron sources, to which we have added  another candidate in SN~1991bg-like events\cite{cro17}, a rare class of supernova.

\begin{table*}
\begin{center}
\caption{Candidate positron sources from Prantzos et al.\cite{pra11}}
\label{tab:cand}       
 \begin{threeparttable}
\begin{tabular}{llll}
\hline\noalign{\smallskip}
Source & Process & Production rate& Bulge/disc  \\
&& ($\times 10^{43}$~e$^{+}$~s$^{-1}$) &\\
\noalign{\smallskip}\hline\noalign{\smallskip}
Massive stars & $^{26}$Al $\beta^{+}$ decay & 0.4 & $<0.2$ \\
Supernov\ae & $^{24}$Ti $\beta^{+}$ decay & 0.3 & $<0.2$ \\
Supernov\ae\ Ia & $^{56}$Ni $\beta^{+}$ decay & 2 & $<0.5$ \\
Supernov\ae\ 1991bg$^{a}$& $^{44}$Ti $\beta^{+}$ decay & 5 & $\sim0.4$ \\
Nov\ae & $\beta^{+}$ decay & 0.02 & $<0.5$ \\
Hypernov\ae/ GRB  & $^{56}$Ni $\beta^{+}$ decay & ? & $<0.2$ \\
Cosmic rays & $p-p \rightarrow \pi^{+}$ & 0.1 &$<0.2$\\
LMXRBs & $\gamma-\gamma$ & 2 & $<0.5$ \\
Micro-quasars & $\gamma-\gamma$ & 1 & $<0.5$ \\
Pulsars & $\gamma-\gamma$/$\gamma-\gamma_{\rm B}$ & 0.5 & $<0.2$ \\
ms pulsars& $\gamma-\gamma$/$\gamma-\gamma_{\rm B}$ & 0.15 & $<0.5$ \\
Magnetars & $\gamma-\gamma$/$\gamma-\gamma_{\rm B}$ & 0.16 & $<0.2$ \\
SMBH &  $p-p \rightarrow \pi^{+}$/ $\gamma-\gamma$ & ? &\\
\noalign{\smallskip}\hline
\end{tabular}
\begin{tablenotes}
      \item$^{a}$From Crocker et al.\ (2017)\cite{cro17}.
 \end{tablenotes}
 \end{threeparttable}
\end{center}
\end{table*}

The candidates in Table~\ref{tab:cand} provide too many photons to account for the total rate of $\approx 5 \times 10^{43}$~e$^{+}$~s$^{-1}$.  However, it should be noted that the rates of production are very uncertain except for massive stars, SN $^{24}$Ti and cosmic rays, and the other values should be taken as upper limits.  With the exception of SN 1991bg-like events all sources have a bulge/disc ratio lower than the observed 511~keV emission; SN 1991bg like events are expected to follow the distribution of old stellar populations, and will therefore have a bulge/disc ratio of $\sim 0.4$, however the rate of these events is currently unknown, but  should in principle be measurable from the abundance of $^{44}$Ca produced by these events\cite{pra17}.
Note that if e$^{+}$ do not propagate far from their sources then positrons from an old stellar population can explain the small offset of the bulge emission to negative Galactic longitudes, which arises naturally from the projection of an inclined long bar\cite{ali14}, such as that seen in the Milky Way\cite{bland16}.  Note that attempts to explain the offset due to young stars in the disc\cite{hig09} are challenged by the lack of a corresponding asymmetry in the $^{26}$Al 1.8 MeV emission\cite{pra11}.

To make a proper assessment of the relative importance of the sources in Table~\ref{tab:cand} one must take into account the location of the sources, the injection energy of the e$^{+}$, the subsequent propagation of the e$^{+}$ and the location of their eventual annihilation.   Despite some laudable  attempts to take on such a challenge\cite{pra06,hig09,ale14}, this task is made difficult due to the lack of detailed knowledge of the properties of the ISM, and especially the magnetic fields, in the inner Galaxy, upon which the propagation depends.  We again refer the reader to Prantzos et al.\ (2011) for a more thorough discussion.

\subsection{Dark matter}

It is often stated that the bulge/disc ratio of the 511~keV emission is well matched by a Navarro, Frenk, and White (NFW) \cite{nav97} dark matter profile\cite{ski14},  
but this ignores the great deal of progress -- discussed below -- on the {\it 
in toto} reconstruction of the bulge's dynamics\cite{bland16}.
 It is also claimed that many classes of dark matter will produce leptons through annihilation, e.g.,
\begin{equation}
\label{eqn:dmann}
\chi + \chi  \rightarrow  {\rm e}^{-} + {\rm e}^{+},
\end{equation}
or decay. Thus some believe that dark matter could provide an explanation of the 511~keV emission.  However, unless the dark matter particles are either very light ($\sim$~MeV) then annihilations of the type in equation~\ref{eqn:dmann} will also produce a continuum of gamma-ray emission\cite{bea05}, $\chi + \chi  \rightarrow  {\rm e}^{-} + {\rm e}^{+} + \gamma$.   Such light dark matter particles are inconsistent with constraints from cosmic microwave background and big bang nucleosynthesis\cite{wil16}.

As is well known to astronomers,
an NFW dark matter profile is {\it completely ruled out} in the Milky Way or 
for that matter any other big galaxy, 
at least within the optical extent. 
The predicted steep rise (`cusp') at the centre is washed out into a flattened `core'. Baryons dominate out to the 
Sun's radius, as is
clearly seen in a recent major review of
the Milky Way\cite{bland16}.
An examination of tens of thousands of kinematic, microlensing and photometric data shows that the baryon fraction of the inner Galaxy is $\approx 0.9$\cite{bland16}, an order of magnitude higher than expected from a NFW profile.
We stress that this core structure in the Galaxy does
not challenge the CDM paradigm because baryon processes in the early universe can effectively flatten the expected cusp\cite{pon12}.

Of course, since it is far from clear what form dark matter actually takes, it is not possible to rule it out.  For example, if dark matter is very heavy, but can exist in different excitation states\cite{fin07} which differ by a few MeV, or else a heavy particle decays to a less massive particle\cite{bou12}, e.g.,
\begin{eqnarray}
\chi^{*} & \rightarrow &  \chi  +  {\rm e}^{-} + {\rm e}^{+}, \\
X & \rightarrow &  \chi  +  {\rm e}^{-} + {\rm e}^{+},
\end{eqnarray}
then it may still be possible to produce the required distribution of e$^{+}$.

However, a recent analysis of the 511~keV emission from the satellite galaxies of the Milky Way suggests that Ps annihilation is not correlated with  dark matter content\cite{sie16c}.   The relative dark matter content of galaxies can be inferred from the ratio of dynamical mass to visible light: the former measures the total mass content of the galaxy, whereas the latter is correlated with the stellar mass content.  A larger mass-to-light ratio implies that a larger fraction of the mass is dark.   It is well established that dwarf galaxies have a larger mass-to-light ratio than more massive galaxies, and are hence more dark matter dominated\cite{mat98,str08,mcc12}.  This trend is expected, since lower mass dark matter haloes are both less able to capture baryons following the reionisation of the ISM  in the early Universe\cite{ree86,tho96}, and less able to retain their baryons against supernov\ae\ driven galactic winds\cite{dek86,kly99}.

If the 511~keV annihilation were due to dark matter annihilation, then $L_{511} \propto M_{\chi}^{2}$, where $M_{\chi}$ is the mass of dark matter particles.  Therefore, for galaxies dominated by dark matter, i.e., $M_{\rm dyn} \sim M_{\chi}$, where $M_{\rm dyn}$ is the dynamical mass, $M_{\rm dyn}/L_{511} \propto 1/M_{\rm dyn}$.  The trend should be the same as for the mass to visible light ratio, though for different reasons.
However, for Ps annihilation the opposite trend is found:
 the mass to 511~keV luminosity ratio  is larger for more massive galaxies\cite{sie16c}, which is inconsistent with an annihilating dark matter origin for the positrons, even after taking into account the low significance of the 511~keV detections for dwarf galaxies.

\section{Recombination lines}
\label{sec:recomb}

The previous sections have summarised the current state of play regarding our knowledge of positronium in the galaxy.  We now turn our attention to the main theme of this paper: the possibility of detecting other signatures of positronium, and the potential information that may result.  

We begin with the possibility of detecting recombination lines of Ps. In fact this was one of the primary motivations of the seminal paper by Mohorovi{\v c}i{\'c}\cite{moh34} in which Ps was first predicted.  However, since positrons annihilate with electrons it was not obvious \emph{a priori} whether Ps will in fact form in astrophysical environments, and moreover in which particular excited states Ps formation can take place.  In fact, $\approx 100\%$ of positrons annihilate via Ps formation (see \S~\ref{sec:hist}, and the discussion around eqn~\ref{eqn:fps}\cite{pra11}).  Note that the calculation  of Ps recombination spectra  must take into account annihilations both before and after formation.

\begin{figure}
\subfigure[]{
\centering \includegraphics[scale=0.6]{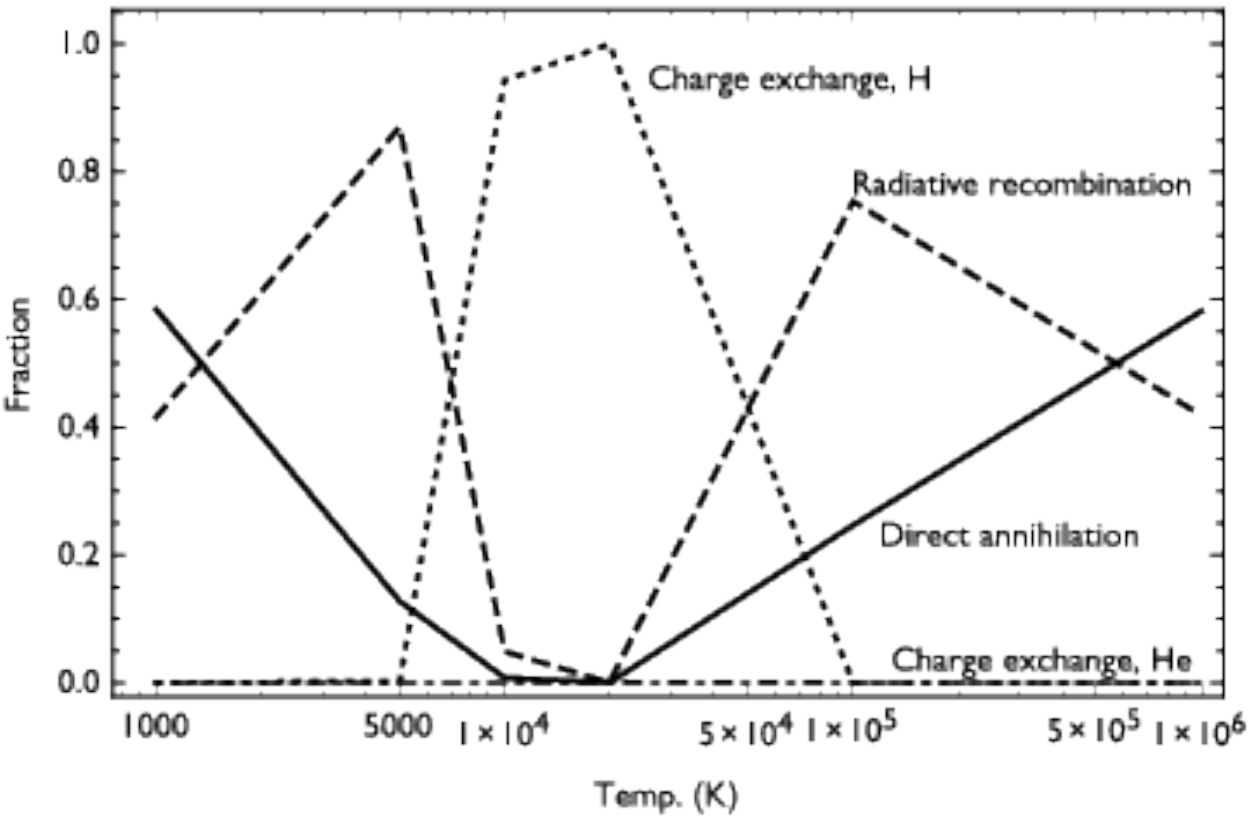}}
\subfigure[]{
\centering \includegraphics[scale=0.6]{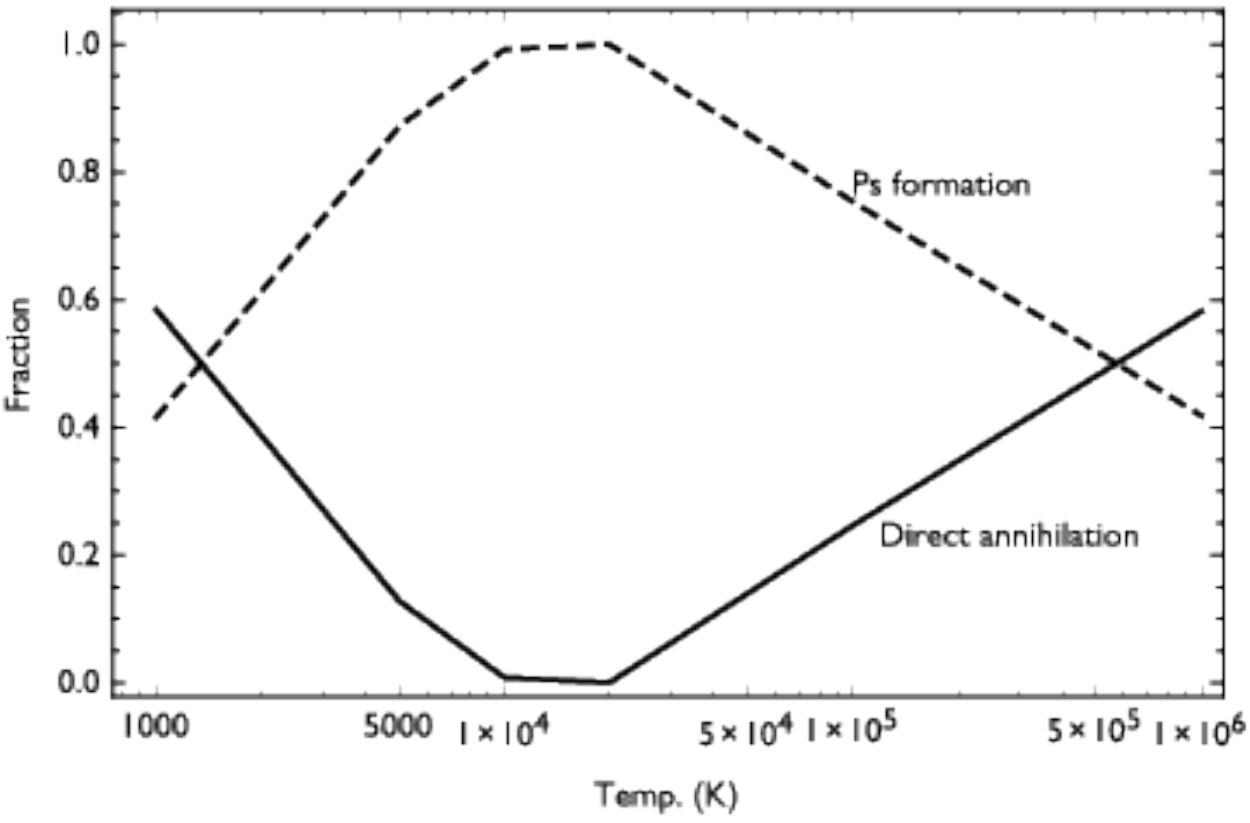}}
\caption{(Top panel.) The fraction of positrons which directly annihilate, or form Ps via radiative recombination, charge exchange with H, or charge exchange with He as a function of temperature.  (Top panel.) The fraction of positrons which directly annihilate, or form Ps via any channel as a function of temperature.  Data are from Wallyn et al. (1996)\cite{wal96}.}
\label{fig:rrdah}
\end{figure}

Positrons in the ISM may either directly annihilate with free or bound electrons, may form Ps through charge exchange with H or He, or may form Ps through radiative recombination.  In a thermal plasma, Ps formation dominates over direct annihilation at temperatures below $6.8 \times 10^{5}$~K\cite{gou89}; in partially ionised media, then charge exhange with H can also play a significant role\cite{wal96}; see Figure~\ref{fig:rrdah}.  In typical ISM conditions Ps formation dominates, and indeed the Ps fraction as measured from the annihilation spectrum is $\approx 1$ within errors\cite{sie16}. 

\begin{figure}
\centering \includegraphics[scale=0.6]{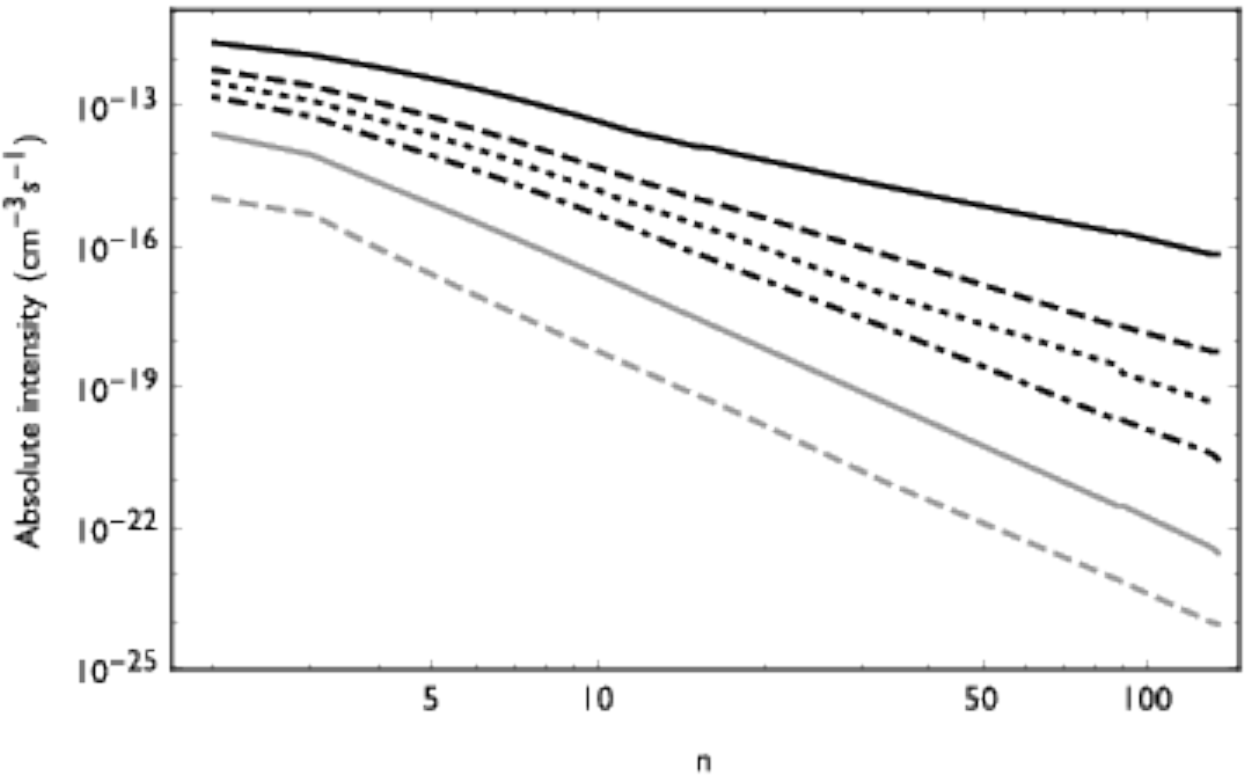}
\caption{The absolute intensities for transitions $n \rightarrow n-1$ as a function of $n$ for various temperatures: 1000~K (black), 5000~K (dashed), 10,000~K (dotted), 20,000~K (dot-dashed), 100,000~K (grey), 1,000,000~K (grey dashed), from table 6 of Wallyn et al.\cite{wal96}.}
\label{fig:nrr}
\end{figure}

Note that the cross-section for radiative recombination is higher for lower quantum levels, and charge exchange is very unlikely in quantum levels $n>2$\cite{wal96}.  Figure~\ref{fig:nrr} shows the absolute intensities of various recombination lines at select temperatures from table~6 of Wallyn et al. (1996)\cite{wal96}, which also take into account the probabilities of annihilation from each state.  

Ceteris paribus, the strongest recombination signatures would be those of the Lyman series, then the Balmer series etc.   However, the extinction due to Rayleigh scattering by dust in the interstellar medium is proportional to $\lambda^{-4}$, and therefore affects the shorter wavelength lines much more than the longer wavelength lines; the wavelengths of the main transition lines are given in Table~\ref{tab:lambda}, and are almost exactly twice those of the equivalent hydrogen lines since the reduced mass for Ps is,
$\mu = \frac{m_{\rm e}}{2}$, and for H it is $\mu_{\rm H} = (m_{\rm e} m_{\rm p})/(m_{\rm e} + m_{\rm p}) \approx m_{\rm e}$.  For typical visual extinctions towards the centre of the Galaxy of $1.5 \le A_{\rm V} \le 5$ Balmer $\alpha$ in the NIR will be the brightest line, for regions of very high extinction, $A_{\rm V} \sim 25$ Paschen $\alpha$ will be stronger, whilst for extremely high extinctions of $A_{\rm V} \sim 50$ Brackett $\alpha$ will be strongest\cite{wal96}.

\begin{table}
\caption{Wavelengths of  select recombination lines of Ps.}
\label{tab:lambda}
\begin{tabular}{ll}
$n \rightarrow n'$ & $\lambda$ ($\mu$m) \\ \hline
 $2 \rightarrow  1 $ (Lyman $\alpha$) & 0.243005 \\
 $3 \rightarrow  1 $ (Lyman $\beta$)& 0.205035 \\
 $3 \rightarrow  2 $ (Balmer $\alpha$)& 1.31222 \\
 $4 \rightarrow  2 $ (Balmer $\beta$)& 0.97202 \\
 $4 \rightarrow  3 $ (Paschen $\alpha$)& 3.7492 \\
 $5 \rightarrow  3 $ (Paschen $\beta$)& 2.5629 \\
 $5 \rightarrow  4 $ (Brackett $\alpha$)& 8.1002 \\
 $6 \rightarrow  4 $ (Brackett $\beta$)& 5.2489 \\
 $6 \rightarrow  5 $ (Pfund $\alpha$)& 14.912 \\
 $7 \rightarrow  5 $ (Pfund $\beta$)& 9.3025 \\
 $7 \rightarrow  6 $ (Humphreys $\alpha$)& 24.730 \\
 $8 \rightarrow  6 $  (Humphreys $\beta$)& 14.997 \\
 $8 \rightarrow  7 $& 38.103 \\
 $9 \rightarrow  7 $& 22.605 \\
 $9 \rightarrow  8 $& 55.577 \\
 $10 \rightarrow  8 $& 32.401 \\
 $10 \rightarrow  9 $& 77.70 \\
 $11 \rightarrow  10 $& 105.01 \\
 $12 \rightarrow  11 $& 138.07 \\
 $13 \rightarrow  12 $& 177.41 \\
 $14 \rightarrow  13 $& 223.59 \\
 $15 \rightarrow  14 $& 277.15 \\
 $16 \rightarrow  15 $& 338.6 \\
 $17 \rightarrow  16 $& 408.6 \\
 $18 \rightarrow  17 $& 487.6 \\
 $19 \rightarrow  18 $& 576.1 \\
 $33 \rightarrow  32 $& 3127. \\
 $88 \rightarrow  87 $& 61000. \\
 $90 \rightarrow  89 $& 65300. \\
 $130 \rightarrow  129 $& 197900. \\
 $134 \rightarrow  133 $& 216800. 
 \end{tabular}
 \end{table}

In practice, the best observing strategy must take into account the expected e$^{+}$ production rates of typical sources, their location in the Galaxy, the likely e$^{+}$ propagation, the expected background emission and the sensitivity of the instruments.  The first attempt to quantify the expected Ps recombination line strengths was made by McClintock (1984)\cite{mcc84}, who was motivated by the planned space telescope, which was eventually to become the Hubble Space Telescope, and found that Lyman $\alpha$ emission from the Crab pulsar and NGC~4151 should be detectable in principle.  The issue of detecting Ps recombination lines has subsequently been revisited several times\cite{burd92,burd96,wal96,burd97}, and most recently by the authors\cite{ell09} based on recent advances in near infrared spectroscopy.

So far, however, Ps recombination lines from astrophysical sources have escaped detection.  There have been three published attempts.  The first\cite{ana89} searched for the 87$\alpha$ radio recombination line, under the erroneous assumption that all Ps atoms would form at high $n$ levels and cascade to the ground-state before annihilation (this work preceeded the detailed calculations of the recombination spectrum of Ps\cite{wal96}).  In fact, as discussed above, the radio recombination lines will be rather improbable, see Figure~\ref{fig:nrr}.  This conclusion is supported by further recent attempts to measure the Ps131$\alpha$, Ps132$\alpha$, Ps133$\alpha$ and Ps135$\alpha$ radio recombination lines from the Galactic centre\cite{rey17}, which find upper limits to recombination rate of $< 3.0 \times 10^{45}$~s$^{-1}$.
Puxley and Skinner\cite{pux96} searched for Paschen $\gamma$ from the Galactic centre.  They found an upper limit of  upper-limit to the line strength of $3 \times 10^{-19}$~W~m$^{-2}$, which is not constraining compared to the expected line strength\cite{ell09} due to the intrinsically faint line strength, the high NIR background and the relatively  poor sensitiviy of NIR detectors at that time.  

The non-detection of Ps recombination lines is not surprising when detailed calculations expected emission line strengths \emph{and} the astronomical backgrounds and instrument sensitivities are made\cite{ell09}.  In particular the Ps Balmer $\alpha$ line is very close in wavelength to a bright atmospheric OH line.  Recent developments in near-infrared spectroscopy, which suppress these lines\cite{ell08,bland11b,ell12a,tri13a}, or space-based observatories\cite{gar06}, should make it possible to detect these lines in the future.  Doing so would yield an improvement in angular resolution of a factor $\sim 18,000$, which would make it possible to resolve and detect Ps point sources, if any such exist, or  in any case to identify more accurately 
the locations in which Ps forms.

\subsection{Fine and hyperfine transitions}
\label{sec:spin}

\begin{figure}
\begin{center}
\subfigure[Triplet state]{
\includegraphics[scale=0.23]{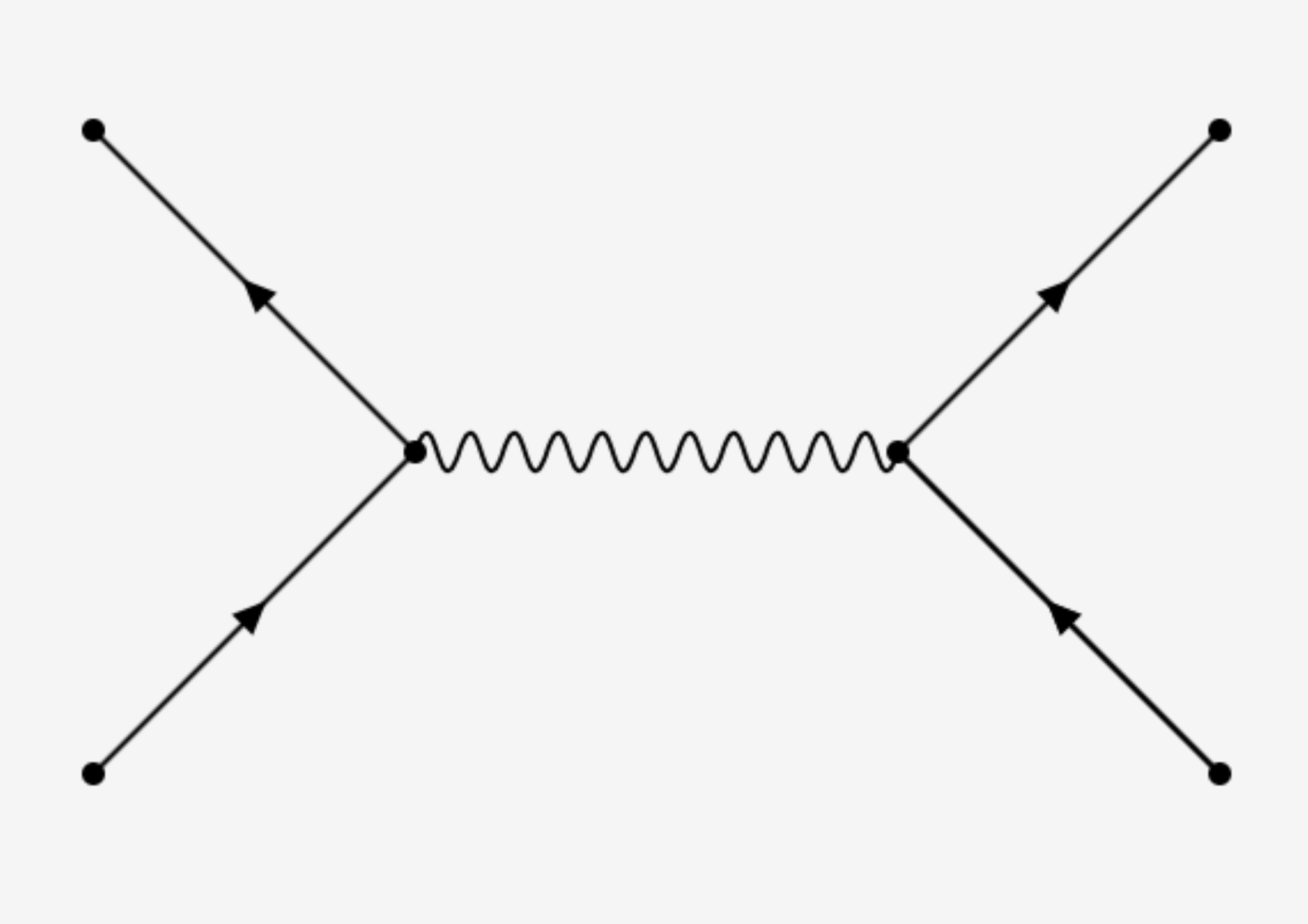}
}
\subfigure[Singlet state]{
\includegraphics[scale=0.27]{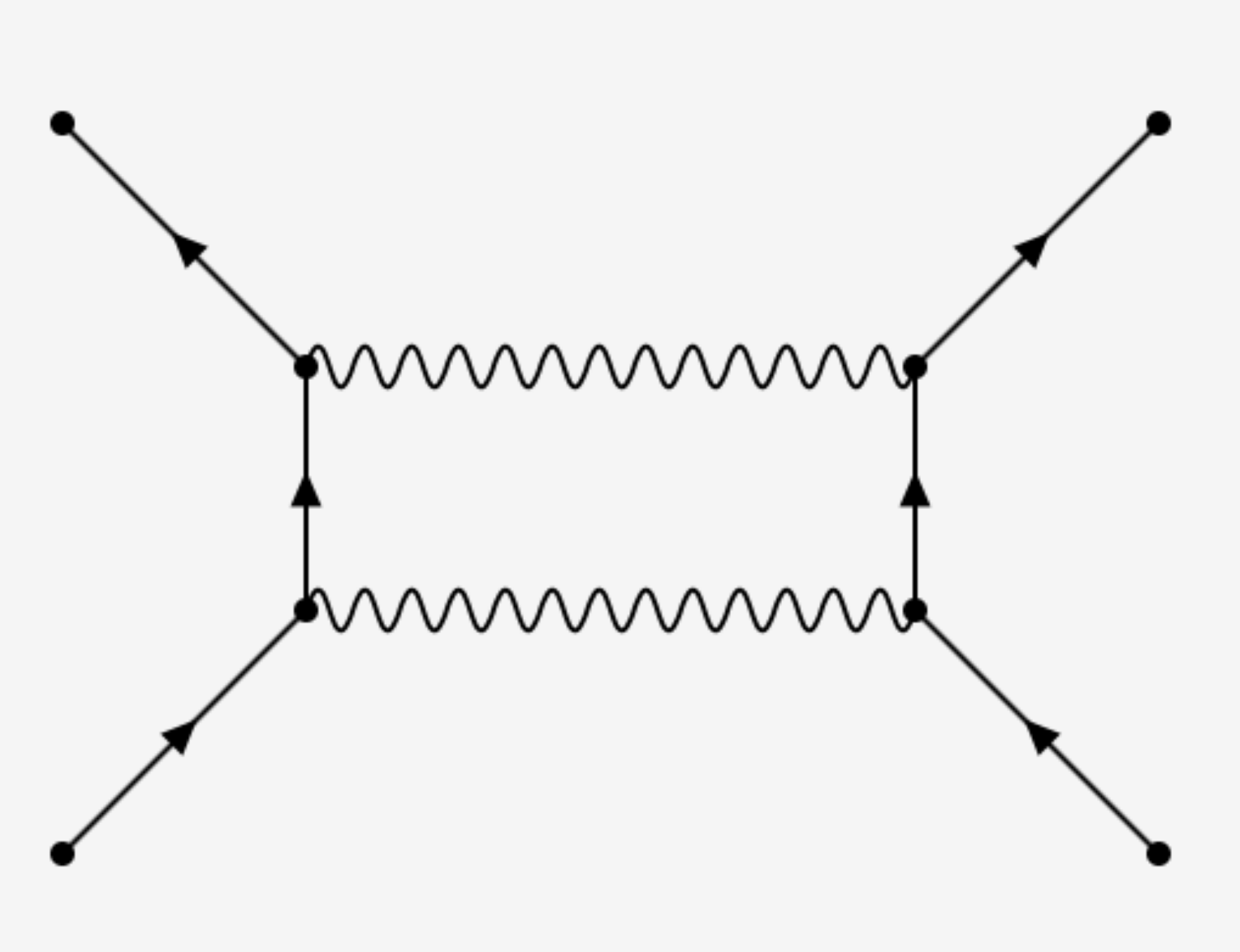}
}
\caption{Examples of two of the lowest order corrections to the energy levels of Ps, due to pair annihilation and creation.  The triplet state annihilation into a single virtual photon introduces a correction of order $m \alpha^{2}$, whereas the singlet state annihilation into two virtual photons introduces corrections of order $m \alpha^{3}$.}
\label{fig:feyn}
\end{center}
\end{figure}

\begin{figure*}
\centering \includegraphics[scale=0.5]{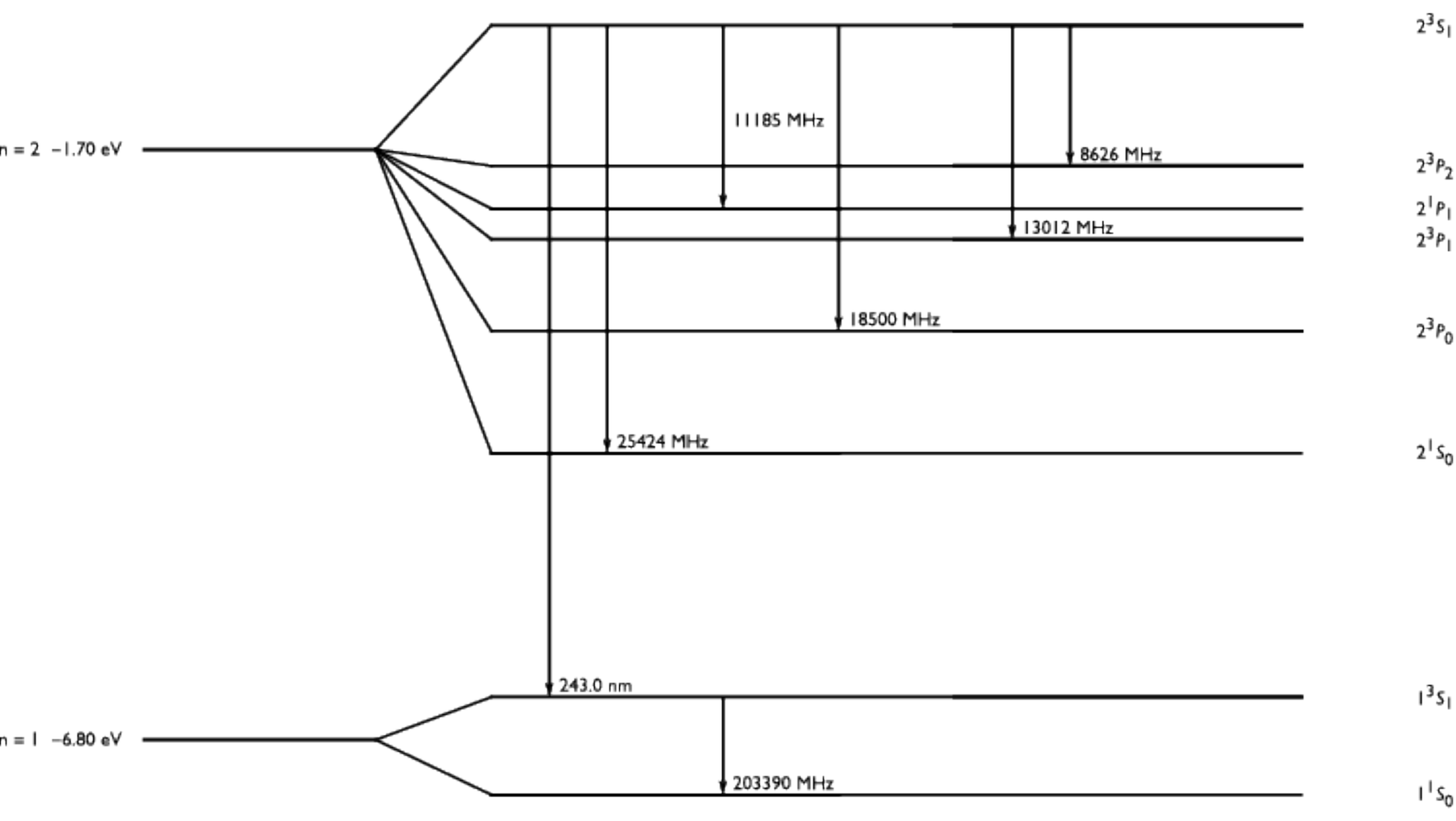}
\caption{Ps energy level fine and hyperfine splitting.}
\label{fig:pslevels}
\end{figure*}

The Ps energy levels will display fine and hyperfine splitting as for hydrogen.  However, the energy levels are modified due to the possibility of pair annihilation and creation\cite{bet57}, as shown in Figure~\ref{fig:feyn} for two of the lowest order corrections.  To order $m \alpha^{2}$, the wavelength of any energy level, $n,l,S,J$ is given by\cite{bet57},
\begin{equation}
E_{nlSJ} = -\frac{\rm Ry}{2} +\left(\frac{11}{32 n^{4}} + \left( \epsilon_{lSJ} - \frac{1}{2l+1}\right)\frac{1}{n^{3}}\right) \alpha^{2} {\rm Ry},
\end{equation}
where
\begin{eqnarray}
\epsilon_{l,S=0,J} &= &0,\\
\epsilon_{l,S=1,J} &=& \frac{7}{6} \delta_{l0} + \frac{1-\delta_{l0}}{2(2l+1)}
\begin{cases}
\frac{3l+4}{(l+1)(2l+3)}, & {\rm  if}\ J=l+1 \\
-\frac{1}{l(l+1)}, & {\rm if}\ J=l\\
-\frac{3l-1}{l(2l-1)}, &  {\rm if}\ J=l-1
\end{cases}
\end{eqnarray}
and Ry is the Rydberg energy of the ground state of \emph{hydrogen},
\begin{equation}
{\rm Ry} = \frac{m_{\rm e} e^{4}}{2 \hbar^{2} 4 \pi \epsilon_{0}^{2}},
\end{equation}
and thus $Ry/2$ is the ground state energy of Ps, $\delta_{l0}$ is the Kroenecker delta function, $\alpha$ is the fine structure constant, $m_{\rm e}$ is the electron mass, and $e$ is the electron charge.  For the case of Ps, the energy splitting due fine and hyperfine transitions are of the same order, unlike for hydrogen in which case the fine structure splitting is greater by a factor of the order $m_{\rm p}/m_{\rm e} \approx 1836$ where $m_{\rm p}$ is the proton mass.  Furthermore, $\approx 3/4$ of the  hyperfine energy splitting is due to the creation and annihilation of pairs\cite{gri82}.
Calculations for the singlet state have been carried out up to order $m \alpha^{7}$\cite{adk16}.  Here we adopt the current highest precision laboratory measurements of the fine\cite{mil75b,hat87,hag93,ley02} and hyperfine\cite{mil75,mil83,rit84,ish14} levels, as shown in Figure~\ref{fig:pslevels}.

The relative population of these levels will in principle allow transitions between the states\cite{burd99b}.  However, the probabilities of such transitions are low, since the lifetime of the states are much shorter than the radiative lifetimes of the transitions\cite{pra11}, see Table~\ref{tab:pslife}.  Nevertheless, some small fraction of Ps may undergo such transitions, especially in high density environments such as stellar interiors, when collisionally induced spin-flip transitions will become possible\cite{mur05}.

\begin{table}
\caption{Lifetimes for fine and hyperfine transitions compared to the annihilation lifetimes.}
\label{tab:pslife}
\begin{tabular}{lll}
State& Transition& Lifetime (s) \\ \hline 
 $1^{3}S_{0}$ &Annihilation& $1.41 \times 10^{-7}$ \\ 
&$\rightarrow 1^{1}S_{0}$ spin-flip (hfs) & $2.97 \times 10^{7}$ \\
 $2^{3}S_{1}$ & Annihilation&$1.1 \times 10^{-6}$ \\
&$\rightarrow 2^{3}P_{2}$ fs & $3.46\times 10^{5}$\\
&$\rightarrow 2^{3}P_{1}$ fs & $1.68\times 10^{5}$\\
&$\rightarrow 2^{3}P_{0}$ fs & $1.75\times 10^{5}$\\
\end{tabular}
\end{table}

Note that $l=1$ cannot annihilate.

\section{Leptonium}
\label{sec:lep}

An interesting possibility is the formation of the analogous leptonic atoms, true muonium $\mu^{-} - \mu^{+}$ and true tauonium $\tau^{-} - \tau^{+}$.
Initial considerations\cite{mof75} were dismissed due to the short decay times of the constituent particles\cite{avi79,burd91}.  Recently however, we have revisited the question of leptonium in astrophysical environments\cite{ell15}.  

Because of the  intrinsic instability of the particles to decay, leptonium atoms will only form for a small fraction, $\sim 10^{-6}$ -- $10^{-7}$, of pairs produced with energies less than the ionisation energy of the atoms.  Of these pairs, most will form leptonium, the cross-section for annihilation being much smaller under astrophysical conditions.  
After formation, it is possible to again produce recombination lines before annihilation, though this is complicated by the fact that the constituent particles may decay.  

Nevertheless, there are significant branching ratios for both two photon annihilation and recombination lines for  true muonium and small probabilities for true tauonium.  Furthermore, even improbable transitions may give rise to significant signatures if the lepton pairs are produced in sufficient quantity - a situation which is common in  high energy astrophysical environments.  Table~\ref{tab:lep} summarises the main observational signatures.  In practice, leptonium formation within micro-quasar jet-star interactions, or within the accretion discs of both AGN and micro-quasars could yield signatures brighter than current detection limits, with those from micro-quasars offering the brightest estimates due to their proximity.  However, full calculations of the signal-to-noise taking into account intrinsic backgrounds and  emission from the sources have not been made.

\begin{table}
\caption{The main signatures and branching ratios for true muonium and true tauonium}
\label{tab:lep}
\begin{tabular}{lllll}
\hline \hline 
\multicolumn{5}{c}{True muonium} \\ \hline 
Transition & Energy & \multicolumn{3}{c}{Branching ratio} \\
&&1000~K & 10000~K & 100000~K \\ \hline \hline
$\gamma-\gamma$ &105.66~MeV & 0.16	&0.2	 & 0.23 \\
Ly $\alpha$ & 1.055~keV& 0.37	&0.45 & 0.46 \\
Ly $\beta$ & 1.250~keV& 0.04& 0.05 & 0.07 \\
Ba $\alpha$ &0.195~keV & 0.27&0.31 & 0.26\\
Ly $\beta$ & 0.264~keV& 0.04& 0.05& 0.06\\ 
\hline \hline 
\multicolumn{5}{c}{True tauonium} \\ \hline
Transition & Energy & \multicolumn{3}{c}{Branching ratio} \\
&&1000K & 10000K & 100000K \\ \hline \hline
$\gamma-\gamma$ & 1.4066~GeV & 0.03& 0.03& 0.04\\
Ly $\alpha$ & 17.7~keV& 0.01& 0.01& 0.01
\end{tabular}
\end{table}

\section{Summary}
\label{sec:summ}

Gamma-ray observations over the last 40 years have converged upon a model in which $\approx 5 \times 10^{43}$~e$^{+}$ annihilate every second in our Galaxy.   Almost 100~\%\ of these annihilate via Ps formation in the warm ionised interstellar medium.  The Ps is distributed in two bulge components and a broad disc, along with a compact central source.  There is no evidence for time variability\cite{sie16}.

There are many candidates for the sources of the positrons, amongst the most promising being young massive stars, low mass X-ray binaries, and various types of supernov\ae.  Sophisticated models taking into account the positron production rates, and positron propagation through the interstellar medium are now shedding light on the relative importance of these sources\cite{pra11}.
Recent important observations  have found  evidence for 511~keV radiation from the V404 Cygni micro-quasar, the first unambiguous evidence for a specific source of e$^{+}$\cite{sie16b}; and also an anti-correlation in 511~keV radiation with mass-to-light ratio in the satellite galaxies of the Milky Way\cite{sie16c}, which argues against a dark matter origin for the positrons. 

Despite these considerable achievements, the chief difficulty in furthering our understanding of positron and positronium formation in the Galaxy is the relatively poor angular resolution of gamma-ray telescopes, which is an unavoidable consequence of observing high energy photons, which cannot be focussed.  
We have reviewed the possibility of observing other signatures of Ps, and their potential to provide more information on the sources of Galactic positrons.

The main candidate is the observation of Ps recombination lines.  However, since this idea was first suggested over 80 years ago\cite{moh34}, and despite periodic revivals of interest\cite{mcc84,ana89,burd92,burd96,wal96,pux96,burd97,ell09}, to date there have been no detections of any astrophysical Ps recombination lines.  The main reason for this is the high extinction of the UV Lyman lines, the high background of the NIR Balmer lines, and the intrinsic faintness of the longer wavelength lines\cite{wal96,ell09}.   Recent advances in near-infrared spectroscopy\cite{ell08,bland11b,ell12a,tri13a}, or space-based observatories\cite{gar06}, should make it possible to detect these lines in the future and thereby  make it possible to resolve and detect Ps point sources, and to test models of e$^{+}$ production, propagation and annihilation.  The most promising sources for such observations are micro-quasars, low mass X-ray binaries and the jets of active galactic nuclei (AGN)\cite{ell09}.
Other transitions, such as radio recombination lines and spin-flip\cite{burd99b} are less promising.  Radio recombination lines are unlikely due to the low cross-section for recombination into highly excited states (\S~\ref{sec:recomb}), unless a sufficiently luminous ionising source is able to excite the Ps atoms before annihilation.  Fine and hyperfine transitions are unlikely due to the small transition probabilities compared to the probabilities for annihilation from the same states (\S~\ref{sec:spin}). 

An interesting alternative approach is to look for analogous sources of true muonium and true tauonium\cite{ell15}. The probabilities of forming leptonium are very small, due to the fact that the constituent particles decay, and also that only a small fraction of lepton pairs produced will have sufficiently low energy to recombine into an atom.  Nevertheless, given a sufficiently productive source, significant numbers of leptonium atoms can form.  These will have significant branching fractions into two photon annihilation and Lyman and Balmer recombination lines\cite{ell09}.  The most promising candidates are micro-quasar jet-star interactions, and  the accretion discs of both AGN and micro-quasars.  

Note that because of the intrinsic instability of the $\mu$ and $\tau$ leptons, these particles cannot propagate far through the interstellar medium, so any resulting leptonium must form in situ.     This offers an excellent way to distinguish between a truly diffuse source of leptons (e.g.\ dark matter) and a distribution of unresolved point sources, or a diffuse distribution of $e^{+}$ due to propagation.

%
%
\section{Authors contributions}
All the authors were involved in the preparation of the manuscript.
All the authors have read and approved the final manuscript.
%
\bibliographystyle{epj}
\bibliography{ps}
%
%
%

\end{document}